\documentclass{kluwer} 

\usepackage{graphicx}

\begin{document}



\begin{opening}

\title{Large-Scale Structure in the NIR-Selected MUNICS Survey}

\runningtitle{Large Scale Structure in MUNICS}

\author{C.~S.~Botzler$^1$, J.~Snigula$^1$, R.~Bender$^1$, N.~Drory$^1$, G.~Feulner$^1$,
  G.~J.~Hill$^2$, U.~Hopp$^1$, C.~Maraston$^1$, C.~Mendes de Oliveira$^3$}

\institute{$^1$ University Observatory Munich (USM), Germany \\
$^2$ University of Texas at Austin, USA \\
$^3$ Instituto Astronomico e Geofisico, Sao Paulo, Brazil}




\begin{abstract}
  The Munich Near-IR Cluster Survey (MUNICS) is a wide-area,
  medium-deep, photometric survey selected in the K' band. The
  project's main scientific aims are the identification of galaxy
  clusters up to redshifts of unity and the selection of a large
  sample of field early-type galaxies up to $z < 1.5$ for evolutionary
  studies. We created a Large Scale Structure catalog, using a new
  structure finding technique specialized for photometric datasets,
  that we developed on the basis of a friends-of-friends algorithm.
  We tested the plausibility of the resulting galaxy group and cluster
  catalog with the help of Color-Magnitude Diagrams (CMD), as well as
  a likelihood- and Voronoi-approach.
\end{abstract}

\keywords{galaxies: photometric redshift, survey, cluster finding: friends-of-friends}

\end{opening}


\section{Motivation}

The MUNICS survey was created in order to identify clusters of
galaxies at high redshifts by detecting their luminous early-type
galaxy population and to use the resulting mass function of clusters
for cosmological tests \cite{BFC97}. Another aim of the survey was
to utilize field- and cluster-galaxies over a wide range of redshifts
as a laboratory for galaxy evolution \cite{Drory2001,Feulner2002}.

\section{MUNICS in Brief}

\begin{figure}
  \centerline{
    \begin{tabular}{cc}
      \includegraphics[width=0.49\textwidth]{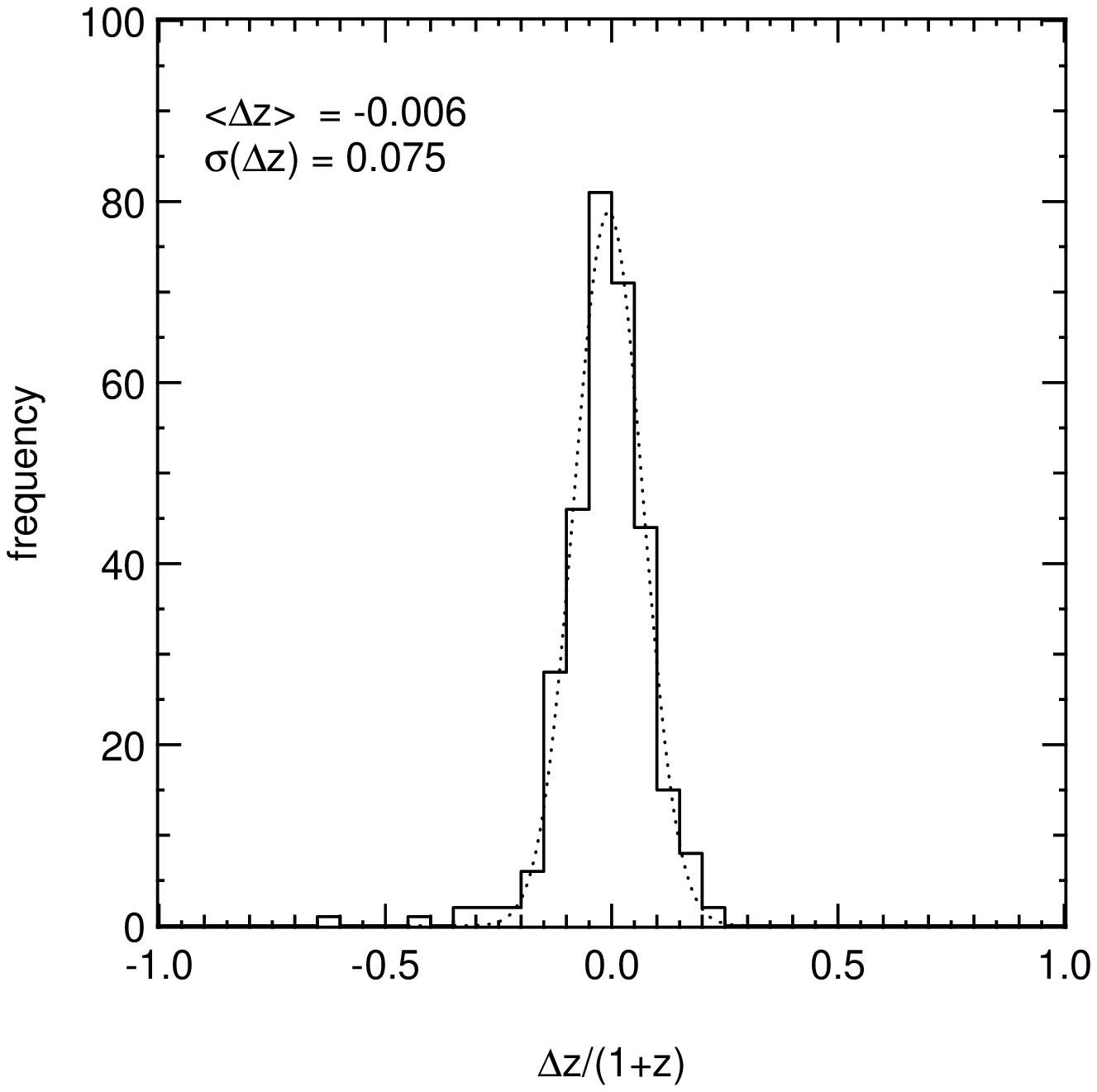} &
      \includegraphics[width=0.49\textwidth]{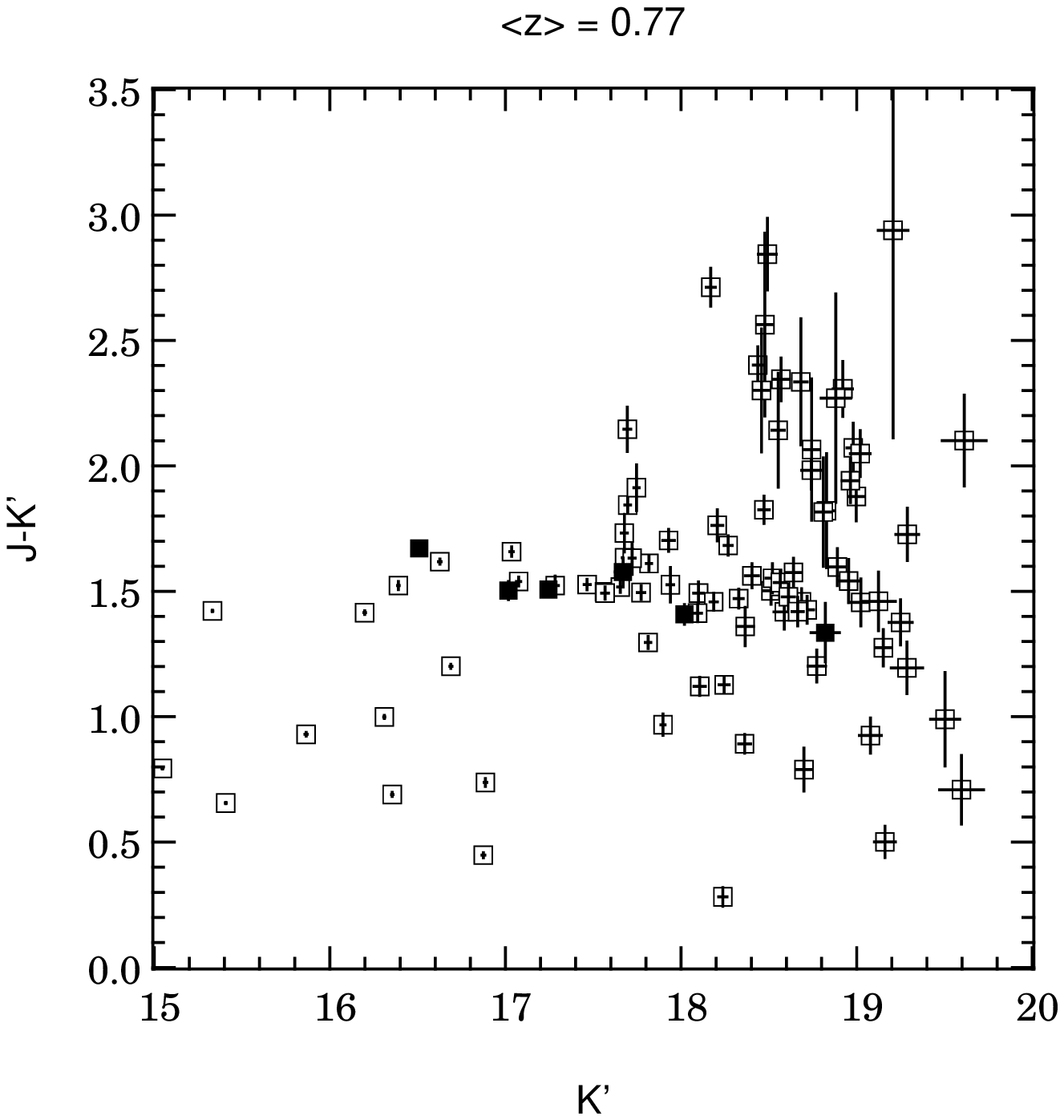} 
    \end{tabular}}
    \caption{\textbf{left panel:} Histogram of the redshift errors.
      The rms scatter is consistent with a Gaussian (dotted line:
      best-fit Gaussian) of a width $\sigma = 0.075$ and an
      insignificant mean deviation from the unity relation of $\langle
      \Delta z \rangle = -0.006$.  \textbf{right panel:} J-K' vs.~K'
      color-magnitude diagram of one group at $\langle z \rangle =
      0.77$. The filled boxes denote the galaxies that were found to
      be structure members by our FOF. The open squares denote all
      galaxies found within 5 core-radii (r$_c$ = 0.125 h$^{-1}_{100}$
      Mpc)(at $\langle z \rangle$) of the cluster center.  The group
      members roughly trace a red-sequence.}
  \label{fig1}
\end{figure}

Our survey covers an area of roughly 1 deg$^2$ in the near-IR J and K'
bands, supplemented with 0.6 deg$^2$ of V, R, and I band imaging.
The galaxy catalog was selected in K' with a 50\% completeness at K'
$\sim$ 19.5$^m$ \cite{MUNICS1}. Redshifts were determined from V, R,
I, J and K' band photometry using a photometric redshift technique
\mbox{\cite{photred00}}, that makes use of model SEDs
\cite{Maraston1998} and empirical templates, and was calibrated with
$\sim$ 500 spectroscopic redshifts \cite{Feulner2002} (see
Fig.~\ref{fig1}, left panel). The catalog contains $\sim$ 5000 galaxies
with $z \geq 0.4$.


\section{The Extended Friends-of-Friends Algorithm}

So far, friends-of-friends (FOF) algorithms \cite{HG1982} were
designed to find number density enhancements in spectroscopic redshift
surveys or numerical simulations \cite{NW87}. Due to the
comparatively large uncertainties in the photometric redshifts, the
original FOF-technique had to be modified for application to MUNICS,
resulting in our ``extended friends-of-friends'' algorithm.

This modified algorithm takes the photometric redshift errors into
account. It looks for friends compatible with an a priori redshift on
a redshift grid, spanning the entire depth of the survey and finally
unifies the structures found in individual redshift bins, if
structures have at least one member in common.

We found 104 structures within $0.4 \leq z \leq 1.1$ in our survey
field, which corresponds to a (flux-limited) number density of $\sim 8
\cdot 10^{-5}$ structures Mpc$^{-3}$ within $0.5 \leq z \leq 0.8$ for
an $\Omega_M = 0.3$, $\Omega_\Lambda = 0.7$, and $h_{100} = 0.7$ cosmology.

\section{Testing the Structures}

\subsection{Color-Magnitude Diagrams}

In a J-K' vs.~K' CMD, evolved members of a bound structure are expected to be
lying roughly on a horizontal line, the ``red sequence'' (see
Fig.~\ref{fig1}, right panel). About 50\% of our CMDs look reasonable for
groups or clusters.


\subsection{Voronoi Tessellation}


The inverse of the area of a Voronoi cell around a galaxy is a measure
for the local density of the galaxy distribution and can be used for
finding accumulations of galaxies and finally clusters
\cite{Ramella2001}. 

\begin{figure}
  \centerline{    
    \begin{tabular}{cc}
      \includegraphics[width=0.49\textwidth]{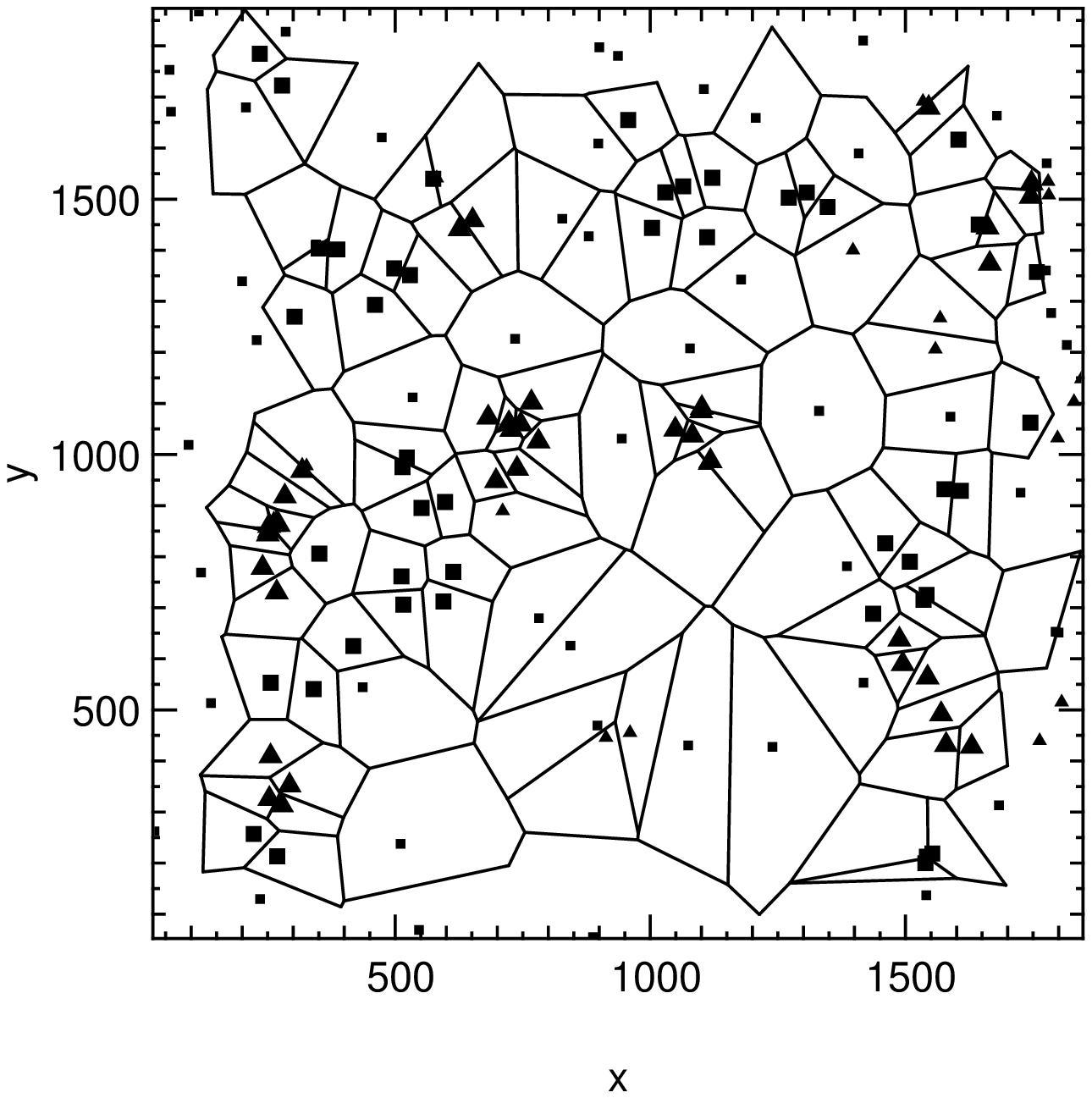} &
      \includegraphics[width=0.49\textwidth]{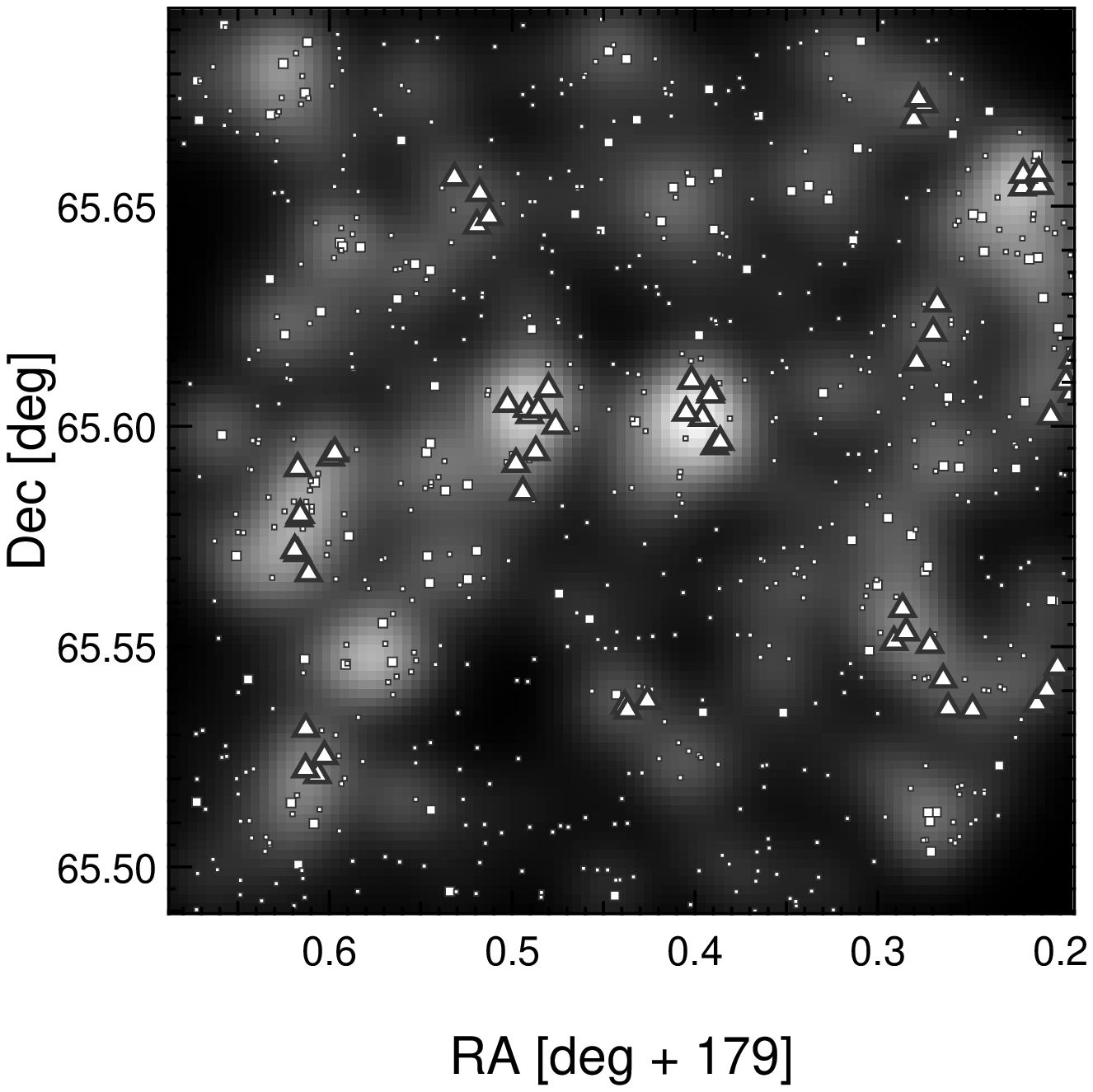} 
    \end{tabular}}
    \caption{\textbf{left panel:} Voronoi tessellations and extended-FOF group members for
      all galaxies with $0.46 \leq z \leq 0.74$ of a MUNICS field.
      Galaxies with Voronoi areas smaller than the mean area of a
      random field Voronoi tessellation (i.e.~overdense) are shown as
      large symbols. Galaxies in underdense regions are denoted by
      small symbols.  Squares denote non-FOF objects. Members of FOF
      structure are plotted as triangles. The graphic shows that they
      lie preferably in overdense regions, suggesting consistency with
      the Voronoi approach.
      \textbf{right panel:} Likelihood distribution and extended-FOF group members
      for all galaxies within $0.45 \leq z \leq 0.75$ of the same
      MUNICS field as seen in Fig.~\ref{fig2}, right panel. The likelihood is
      shown in grey-scales, with white symbolizing the highest
      probability.  Members of our FOF structures are plotted as open
      black triangles. The graphic shows the tendency for the FOF
      structures to lie in areas of increased likelihood.}
    \label{fig2}
\end{figure}

We prepared Voronoi tessellations in preselected photometric redshift
bins for our MUNICS dataset and compared the resulting density charts
with the results of our FOF algorithm (see Fig.~\ref{fig2}, left panel).

\subsection{Likelihood Approach}

We furthermore determined the probability that a given point on a
uniform grid in redshift-space is the center of a galaxy cluster,
taking into account the known distributions of galaxies in the MUNICS
survey. To estimate this likelihood, we assumed Gaussian probability
distributions for a typical cluster in redshift, projected position
and J-K' color. We included empirical knowledge about typical cluster
core radii (r$_c$ = 0.125 h$^{-1}_{100}$ Mpc; \opencite{Bahcall1977}) and
velocity dispersions ($\sigma_{V}$ = 10$^3$ km s$^{-1}$), as well as
the individual errors of the MUNICS galaxies and SSP model colors
\cite{Maraston1998}
to define the rms of the distributions (see Fig.~\ref{fig2}, right panel).

\begin{acknowledgements}
The MUNICS project was suported by the Deutsche Forschungsgemeinschaft,
Sonderforschungsbereich 375.
\end{acknowledgements}






%
  

\enlargethispage{1cm}

\bibliographystyle{klunamed}
\bibliography{Botzler}



\end{document}